\documentstyle[11pt,aaspp4,flushrt]{article}

\def\zg5a{0140+326~RD1}

\def\spose#1{\hbox to 0pt{#1\hss}}
\def\Mdot {\ifmmode {\rm {\dot M}} \else ${\rm {\dot M}}$\fi}

\def\kms{\ifmmode {\rm\,km\,s^{-1}}\else
    ${\rm\,km\,s^{-1}}$\fi}
\def\kmsMpc{\ifmmode {\rm\,km\,s^{-1}\,Mpc^{-1}}\else
    ${\rm\,km\,s^{-1}\,Mpc^{-1}}$\fi}

\def\msun{\ifmmode {\rm\,M_\odot}\else ${\rm\,M_\odot}$\fi}
\def\Msun{\ifmmode {\rm\,M_\odot}\else ${\rm\,M_\odot}$\fi}
\def\lsun{\ifmmode {\rm\,L_\odot}\else ${\rm\,L_\odot}$\fi}
\def\Lsun{\ifmmode {\rm\,L_\odot}\else ${\rm\,L_\odot}$\fi}
\def\rsun{\ifmmode {\rm\,R_\odot}\else ${\rm\,R_\odot}$\fi}
\def\Rsun{\ifmmode {\rm\,R_\odot}\else ${\rm\,R_\odot}$\fi}
\def\lya{\ifmmode {\rm\,Ly\alpha}\else ${\rm\,Ly\alpha}$\fi}

\def\cm{{\rm\,cm}}
\def\cm3{\ifmmode {\rm\,cm^{-3}}\else ${\rm\,cm^{-3}}$\fi}

\def\ergps{\ifmmode {\rm\,erg\,s^{-1}}\else ${\rm\,erg\,s^{-1}}$\fi}
\def\ergpscm2{\ifmmode {\rm\,erg\,s^{-1}\,cm^{-2}}\else
    ${\rm\,erg\,s^{-1}\,cm^{-2}}$\fi}

\def\eg{{e.g.}}
\def\deg{\ifmmode {^{\circ}}\else {$^\circ$}\fi}
\def\degr{\ifmmode {^{\circ}}\else {$^\circ$}\fi}
\def\degs{\ifmmode {^{\circ}}\else {$^\circ$}\fi}

\def\etal{{et al.~}}

\def\h3Mpc{h^{-3}{\rm Mpc}^3}
\def\Ho{\ifmmode {\rm\,H_0}\else ${\rm\,H_0}$\fi}
\def\hnot{\ifmmode {\rm\,H_0}\else ${\rm\,H_0}$\fi}
\def\h0{\ifmmode {\rm\,H_0}\else ${\rm\,H_0}$\fi}
\def\hnotunit{\ifmmode {\rm\,km\,s^{-1}\,Mpc^{-1}}\else
    ${\rm\,km\,s^{-1}\,Mpc^{-1}}$\fi}
\def\qnot{\ifmmode {\rm\,q_0}\else ${\rm q_0}$\fi}
\def\q0{\ifmmode {\rm\,q_0}\else ${\rm q_0}$\fi}
\def\ie{{i.e.}}
\def\mic{\ifmmode {\rm\,\mu m}\else ${\rm \mu m}$\fi}

\def\arcsec{\ifmmode {^{\prime\prime}}\else $^{\prime\prime}$\fi}
\def\asec{\ifmmode {^{\prime\prime}}\else $^{\prime\prime}$\fi}
\def\arcmin{\ifmmode {^{\prime}}\else $^{\prime}$\fi}
\def\amin{\ifmmode {^{\prime}}\else $^{\prime}$\fi}

\def\secper{\ifmmode \rlap.{^{s}}\else $\rlap{.}{^{s}} $\fi}
\def\minper{\ifmmode \rlap.{^{m}}\else $\rlap{.}{^m} $\fi}
\def\magper{\ifmmode \rlap.{^{m}}\else $\rlap{.}{^m} $\fi}
\def\arcsper{\ifmmode \rlap.{^{\prime\prime}}\else
    $\rlap.{^{\prime\prime}}$\fi}
\def\arcsper{\ifmmode \rlap.{^{\prime\prime}}\else
    $\rlap.{^{\prime\prime}}$\fi}
\def\arcmper{\ifmmode \rlap.{^{\prime}}\else
    $\rlap.{^{\prime}}$\fi}
\def\spose#1{\hbox to 0pt{#1\hss}}
\def\simlt{\mathrel{\spose{\lower 3pt\hbox{$\mathchar"218$}}
     \raise 2.0pt\hbox{$\mathchar"13C$}}}
\def\simgt{\mathrel{\spose{\lower 3pt\hbox{$\mathchar"218$}}
     \raise 2.0pt\hbox{$\mathchar"13E$}}}

\def\refindent{\par\noindent\parskip=2pt\hangindent=3pc\hangafter=1 }

\def\aa{{A\&A}}

\def\aj{{AJ}}
\def\apj{{ApJ}}

\def\apjsupp{{ApJS}}
\def\apjs{{ApJS}}

\def\mnras{{MNRAS}}
\def\nature{{Nature}}

\def\pasp{{PASP}}

\def\apjref#1;#2;#3;#4 {\par\pp#1, {#2}, #3, #4 \par}

\tighten

\slugcomment{\it Accepted for Publication in ApJ Letters}

\begin{document}

\title{A Galaxy at $z = 5.34$\altaffilmark{1}}
\author{Arjun Dey\altaffilmark{2,3}, 
Hyron Spinrad\altaffilmark{4}, Daniel Stern\altaffilmark{4}, James R.\ Graham\altaffilmark{4}, and 
Frederic H.\ Chaffee\altaffilmark{5}}

\altaffiltext{1}{Based on observations at the W.\ M.\ Keck Observatory.}
\altaffiltext{2}{Hubble Fellow}
\altaffiltext{3}{Dept. of Physics \& Astronomy, The Johns Hopkins University, Baltimore, MD 21218}
\altaffiltext{4}{Astronomy Dept., 601 Campbell Hall, U. C. Berkeley, Berkeley, CA 94720}
\altaffiltext{5}{W.~M.~Keck Observatory, 65-1120 Mamalahoa Hwy., Kamuela, HI 96743}

\begin{abstract}

We report the discovery of \lya\ emission from a galaxy at $z=5.34$,
the first object at $z>5$ with a spectroscopically confirmed redshift.
The faint continuum emission (${\rm m_{AB}(8000{\rm \AA})\approx 27}$),
relatively small rest-frame equivalent width of the emission line
($W_{\rm Ly\alpha}^{rest}\approx 95$\AA), and limits on the
\ion{N}{5}/\lya\ ratio suggest that this is a star--forming galaxy and
not an AGN. The star--formation rates implied by the UV continuum
emission and the \lya\ emission are (in the absence of dust extinction)
fairly modest ($\sim 6~h_{50}^{-2}\ \Msun~yr^{-1}$ for \qnot=0.5).
The continuum luminosity is similar to that of sub-$L^*_{1500}$ star--forming
galaxies at $z\sim3$, and the width of the \lya\ line yields an upper
limit to the mass of $< 2.6\times 10^{10}\Msun$.  The strong emission
line detected in this low-luminosity galaxy provides hope for the
discovery of higher luminosity primeval galaxies at redshifts $z>5$.

\end{abstract}



\section {Introduction}

Several lines of evidence suggest that the earliest epoch of galaxy
formation lies beyond a redshift ($z$) of 5.  For instance, the
presence of metals (in excess of primordial abundances) in high--$z$
damped \lya\ systems (\eg, Lu \etal 1996), quasars (Hamann 1997) and in
$z\sim 2.5-3.5$ star--forming galaxies (\eg, Steidel \etal 1996,
Lowenthal \etal 1997) require metal creation and dispersal at higher
redshifts.  The tight photometric sequences in both low--$z$ and
intermediate--$z$ clusters also attest to a high formation redshift
($z_F$) for at least the elliptical galaxy population (\eg, Stanford,
Eisenhardt \& Dickinson 1995).  Indeed, some ellipticals at $z\sim 1.5$
are observed to contain evolved stellar populations with ages $\simgt
3.5$~Gyr (\eg, Spinrad \etal 1997), again implying $z_F>5$.  

Contemporary searches for the primeval galaxy population at high $z$
have relied on detecting the luminosity from their young stellar
population, targetting either the bright UV continuum emission which
bears the imprint of intergalactic and interstellar neutral hydrogen
absorption (\eg, Steidel \& Hamilton 1992), or the luminous \lya\ or
H$\alpha$ emission which arise in low-extinction HII regions (\eg,
Pritchet \& Hartwick 1990, Lowenthal \etal 1991, Djorgovski, Thompson
\& Smith 1993, M{\o}ller \& Warren 1993, Cowie \& Hu 1998; Pritchet
1994 and references therein). Some of these surveys have met with
considerable success, and have identified large samples of
star--forming galaxies at $z\sim 2.5-3.5$ and a few \lya\ emitters both
in the field (\eg, Cowie \& Hu 1998, Djorgovski \etal 1996) and at
higher redshifts ($z\sim 4.5$) associated with luminous quasars or
radio galaxies (\eg, Djorgovski \etal 1987, Pascarelle \etal 1996, Hu,
McMahon \& Egami 1996).  These objects are unlikely to be true
protogalaxies, undergoing their first epoch of star--formation. The
$z\sim3$ star--forming galaxies appear similar to starburst galaxies in
the local universe: their spectra do not show the large
\lya\ luminosities thought to be characteristic of protogalaxies (\eg,
Meier 1976), and their interstellar media contain heavy elements and
dust (\eg, Steidel \etal 1996, Meurer \etal 1996).  The \lya\ emitters
are more promising protogalactic candidates, but there has been
a dearth of high signal-to-noise data on their continuum emission.

In this letter, we report the discovery of a \lya --emitting galaxy at
$z=5.34$. This is the first galaxy spectroscopically confirmed to be at
$z>5$.  At $z=5.34$, a $\hnot=50\hnotunit$, $\Omega=1$ ($\Omega=0.2$)
Universe is only 820~Myr (1.56~Gyr) old, corresponding to a lookback
time of 93.7\% (90.6\%) of the age of the Universe (we assume
$\Lambda=0$ throughout this paper). Any object observed at this epoch
must necessarily be in its early youth.  We present our observations of
this galaxy in the following section, our determination of its $z$ in
\S~3, and discuss the galaxy's inferred properties in \S~4.

\section {Observations}

%
%
%
%
%

The serendipitous discovery of this galaxy occurred during a successful
search for $z\sim 4$ galaxies using the `dropout' technique pioneered
by Steidel and collaborators (\eg, Steidel \etal 1996; Madau \etal
1996).  Deep $B$ (40~min.), $R$ (35~min.) and $I$ (70~min.) images of a
field centered on the $z=4.4$ radio galaxy 6C~0140+326 (Rawlings \etal
1996) were obtained at the W.~M.~Keck Observatory during 
1996 and 1997 using the Low Resolution Imaging Spectrometer (LRIS;
Oke \etal 1995).  The $B$, $R$, and $I$ images reach limiting magnitudes
(1$\sigma$ in a 2\arcsec\ diameter aperture) of 28.7, 27.4 and
27.7~mag.\ respectively (Vega-based magnitudes are
used throughout this paper, except when stated otherwise).
Spectra of $B$-band `dropout' galaxies were obtained through 1\arcsper5
wide, 14\arcsec\ to 25\arcsec\ long slitlets using LRIS on the Keck II
telescope on UT 1997 September 10 and 11. These observations
were made in PA=192.7\deg\ (east of north) with the 400 l/mm grating
($\lambda_{blaze}\approx8500$\AA; $\Delta\lambda_{FWHM} \approx $11\AA) and the 300 l/mm
grating ($\lambda_{blaze} \approx 5500$\AA; $\Delta\lambda_{FWHM} \approx $14\AA).  The
results of our successful search for $z > 3.6$ galaxies will be
reported elsewhere.

\begin{figure}
\plotfiddle{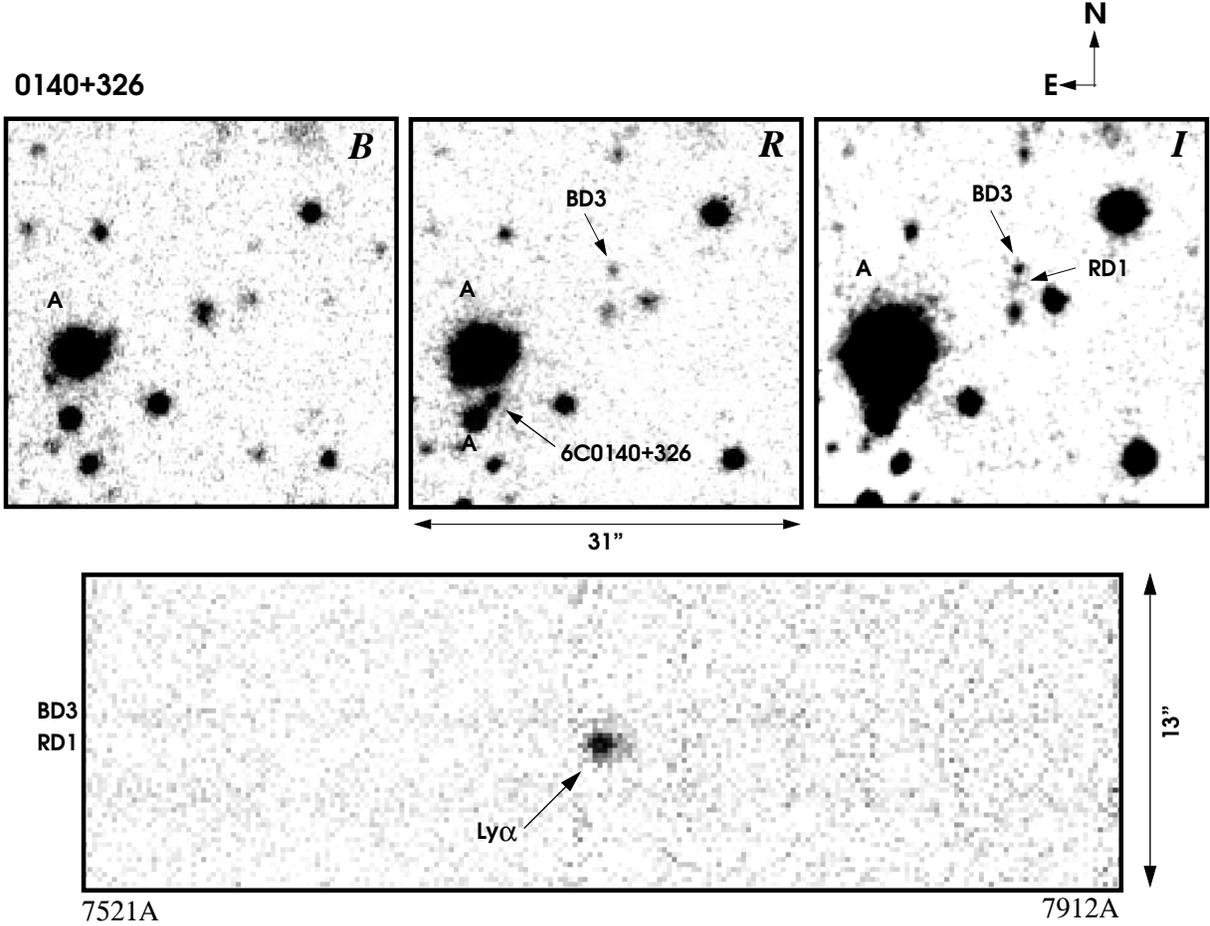}{4.5in}{-90}{60}{60}{-230}{350}
\figcaption{Top: Detail of the Keck $B$, $R$ and $I$ images of the 0140+326
field.  The images shown are 26\arcsec\ on a side, and north is up and
east is to the left.  LRIS has a scale of 0\arcsper21~pixel$^{-1}$, and
the seeing was $\sim 0\arcsper7-0\arcsper9$.  The galaxy originally
targetted as a $B$-band `dropout' is labelled BD3 and lies at $z=4.02$.
The $z=5.34$ galaxy is labelled RD1 in the figure.  Star A is located
at $\alpha=1^h 43^m 43^s.67$, $\delta=+32\deg 53^\prime 54\arcsper3$
(J2000). The offset from star A to RD1 is $\Delta\alpha=-10\arcsper2$
(west) and $\Delta\delta=5\arcsper6$ (north). The \lya\ emission from the
radio galaxy 6C0140+326 ($z=4.41$) is visible near the lower left
corner of the $R$-band image. The object due south of RD1 is an emission 
line galaxy at $z=1.176$. Bottom: A detail of the 2-d spectrum showing 
the Ly$\alpha$ emission line from RD1. The weak continuum from BD3 can be 
seen above the emission line. The data shown here are the coadded 8.5$^h$ 
of spectra obtained using the 400l/mm grating with LRIS.
\label{zg5bri}}
\end{figure}

One targetted $B$--band `dropout' (0140+326~BD3) is a star--forming
galaxy at $z = 4.022.$ The same slitlet also contained a single,
spectrally asymmetric emission line at 7717\AA\ located 1\arcsper2 south
of the continuum emission associated with 0140+326~BD3. The emission
line is associated with  a faint ($I \approx 26.1$ mag in a 1\arcsper5
diameter aperture), spatially resolved (deconvolved FWHM
$\approx$ 0\arcsper7) object in our $I$-band image, but is undetected in
our $B$ and $R$ images to limits of $B>29.0$ and $R>27.8$ (1$\sigma$
in a 1\arcsper5 diameter aperture; figure~\ref{zg5bri}).  We will refer
to this serendipitous $R$-band `dropout' as \zg5a.  We obtained
follow-up spectroscopy of \zg5a\ on UT 1998 Jan. 4 using a
1\arcsec\ wide, $2\arcmper7$ long slit in PA=270\deg\ and 161.2\deg\ with
the 400 l/mm grating to cover $\lambda\lambda$6260--10000\AA, and on UT
1998 Jan. 20 and 21 using a 1\arcsper5 wide, 38\arcsec\ long slitlet in
PA=165.5\deg.

All the reductions were performed using the {\it IRAF} package.
Wavelength calibration was performed using a NeAr lamp; the zero-point
was adjusted, if necessary, using telluric lines.  Flux calibration was
performed using observations of G191B2B, Wolf 1346 and Feige 110
(Massey \etal 1988, Massey \& Gronwall 1990).  Since some of the data
were obtained through cirrus, only observations containing a reliable
detection of the 7717\AA\ emission line were coadded.  The total
spectroscopic integration time on \zg5a\ is 36.2 kiloseconds.  The
spectra were corrected for foreground Galactic extinction using a
reddening of $E_{B-V}\approx 0.035$ (the field lies at a Galactic
latitude of $b^{\rm II}=-28.7\deg$; Burstein \& Heiles 1982).

\section {Results}

\begin{figure}
\plotfiddle{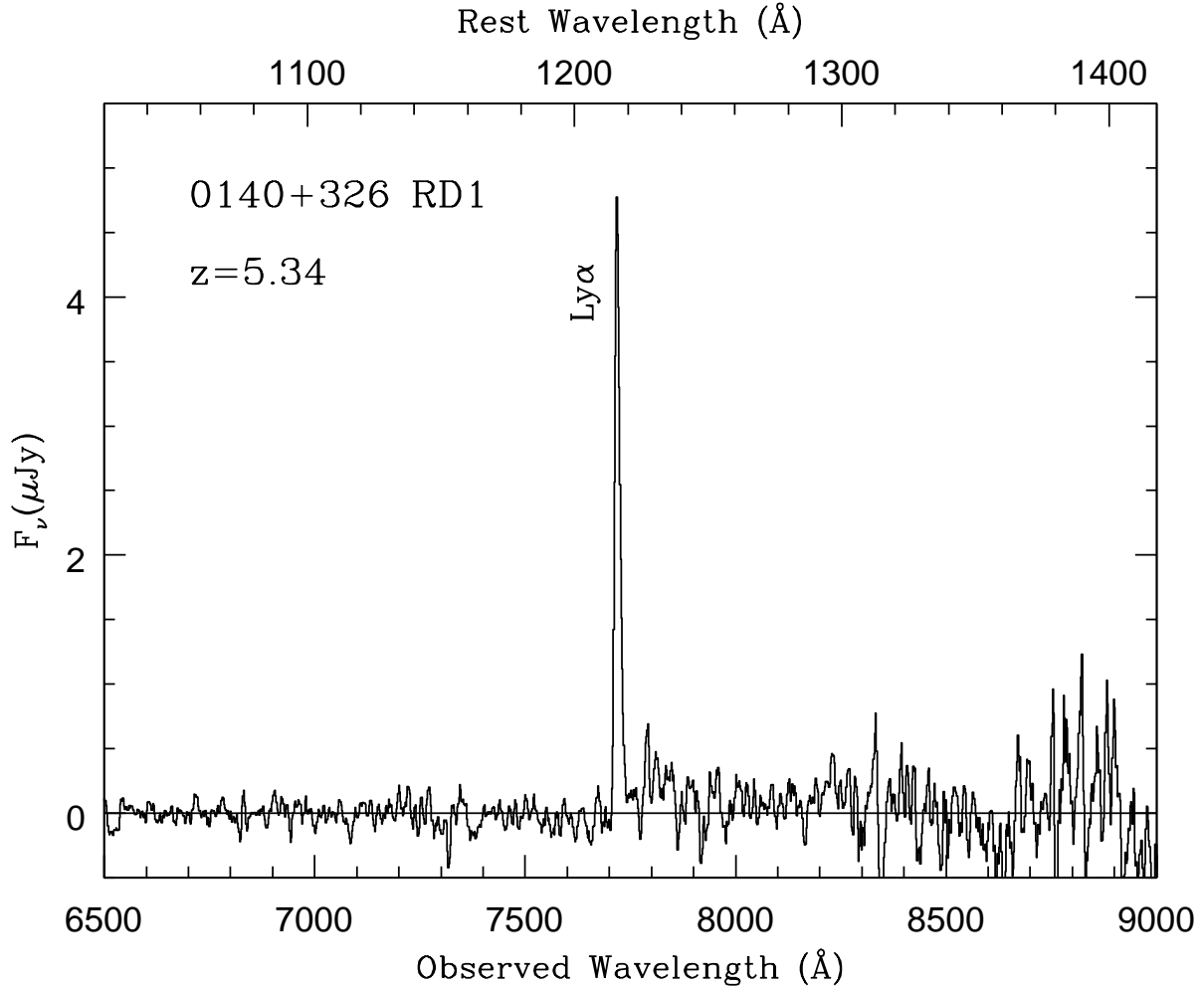}{6in}{0}{100}{100}{-300}{-160}
\figcaption{Coaveraged spectrum of the serendipitously discovered galaxy at 
$z=5.34$. The total exposure time is 36,200 seconds, and the spectrum 
was extracted using an $1\arcsper5\times1\arcsper5$ aperture. The spectrum 
shown has been smoothed using a boxcar filter of width 5 pixels. The 
`features' observed in the continuum are largely due to residuals from
the subtraction of strong telluric OH emission lines (\eg, at 8344\AA).
\label{zg5spec}}
\end{figure}

The spectrum of \zg5a\ shows a strong, spectrally asymmetric emission
line at 7717\AA\ (figure~\ref{zg5spec}).  The asymmetry is observed at
all sampled slit position angles (90\deg, 161.2\deg, 165.5\deg, and
192.7\deg), implying that it is intrinsic to the emission line and not
an artifact of the spatial distribution of the line--emitting gas.  The
spectrum also shows weak continuum emission ($\approx 27$~AB mag for
${\rm 7735\AA<\lambda_{obs}<9000\AA}$) which is observed only redward
of the emission line; we measure an upper limit to the flux blueward of
the line ($\lambda\lambda 6200 - 7700$\AA) of 29.5~AB mag
(1$\sigma$).  Continuum and emission line fluxes for both \zg5a\ and
0140+326~BD3, measured from the dereddened spectrum, are listed in
Table~\ref{tab1}.

At low fluxes ($\sim 1-5\times10^{-17}\ {\rm erg\ s^{-1}\ cm^{-2}}$),
an isolated emission line observed at red optical wavelengths is likely
to be either [\ion{O}{2}]$\lambda\lambda$3726,3729 or \lya; H$\alpha$
or [\ion{O}{3}]$\lambda$5007 can be ruled out by the absence of other
associated emission lines (Balmer H lines,
[\ion{O}{3}]$\lambda\lambda$5007,4959,
[\ion{O}{2}]$\lambda\lambda$3726,3729), the absence of any blue
continuum emission, and (for H$\alpha$) the width and profile of the
observed emission line.  If the emission line at 7717\AA\ were
[\ion{O}{2}] at $z=1.070$, then its rest-frame equivalent width would
be $\approx 300$~\AA, which is in excess of that
produced by even luminous star--forming galaxies (typically less
than 70\AA; Kennicutt 1992, Liu \& Kennicutt 1995).  Moreover,
attempting to fit the observed line profile with an
[\ion{O}{2}]$\lambda\lambda$3726,3729 doublet results in a 3726/3729
ratio of $>$2; this is physically untenable, since the highest ratio
allowed by the statistical weights of the $^2D$ levels is 1.5.

If the observed spatially offset emission line at 7717\AA\ were
associated with the nearby $z=4.022$  galaxy, its most plausible
identity would be \ion{C}{4}$\lambda\lambda$1548,1550 at $z=3.98$,
implying a velocity separation of $\approx 2470~\kms$ from the
\lya\ $z$ of 0140+326~BD3.  However, the complete absence of \lya,
\ion{N}{5}$\lambda\lambda$1239,1243, and
\ion{Si}{4}$\lambda\lambda$1394,1403 emission associated with such a
strong \ion{C}{4} emission system argues against this possibility.  In
addition, the continuum break across
\ion{C}{4}$\lambda\lambda$1548,1550 and the large velocity shift
relative to 0140+326~BD3 would be difficult to explain.

\begin{figure}
\plotfiddle{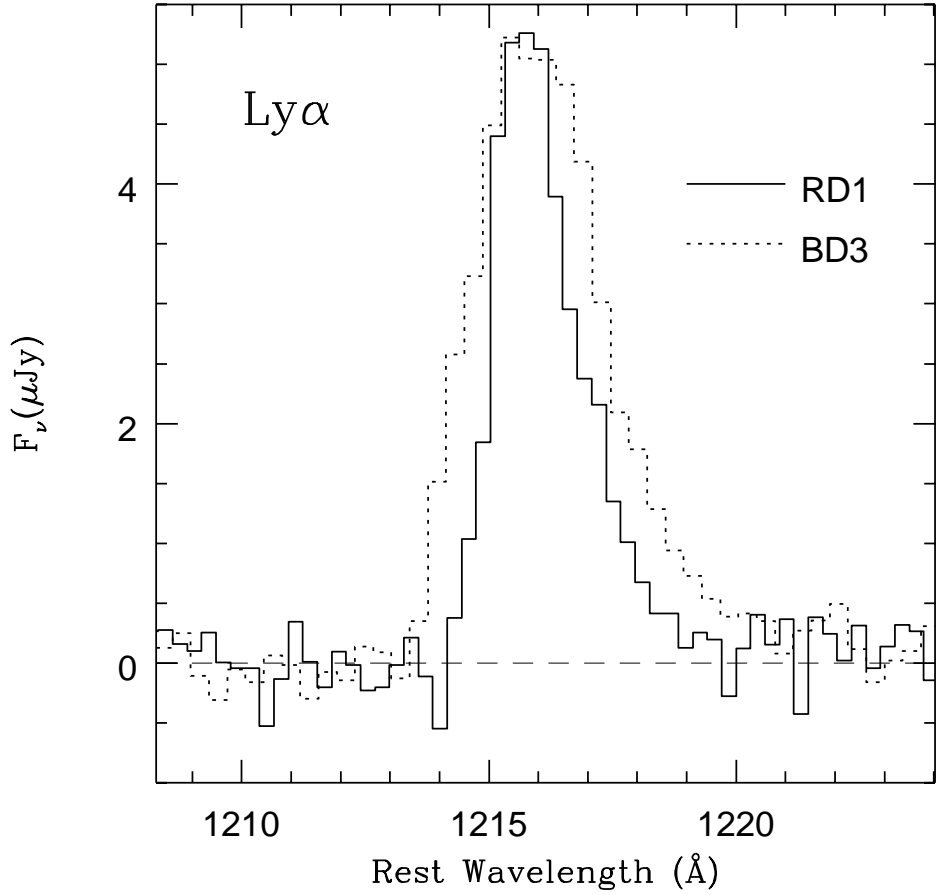}{6in}{0}{100}{100}{-300}{-160}
\figcaption{Detail showing the similarities in the profile asymmetry of
the \lya\ emission line from the $z=5.34$ galaxy \zg5a\ (solid line)
and the $z=4.02$ galaxy 0140+326~BD3 (dotted line).  The spectral
extraction width for \zg5a\ is the same as in figure 2, but the data 
shown here are unsmoothed.
\label{zg5lya}}
\end{figure}

The most plausible interpretation of the 7717\AA\ line is that it is
\lya\ emission from a galaxy at $z=5.34$.  This is supported by the
observed asymmetry of the emission line (\ie, a sharp blueward cutoff
and a red wing) and the continuum break across it. The asymmetry is
typical of high--$z$ \lya\ lines: it is observed, for example, in
high--$z$ radio galaxies, the $z=4.022$ galaxy 0140+326~BD3
(figure~\ref{zg5lya}) and in the $z=4.92$ gravitationally--lensed
galaxy reported by Franx \etal (1997). It is caused largely by
absorbing gas within the galaxy, but with some contribution from
foreground absorbing systems. A similar \lya\ profile is also observed
in local starburst galaxies with galaxian winds (\eg, Lequeux \etal
1995), suggesting that the asymmetry may reflect the presence of a
wind.  The continuum discontinuity at the Ly$\alpha$ line is caused by
the integrated \lya\ absorption due to foreground intervening systems.
The discontinuity is usually described by the broad-band
``flux-deficit'' parameter $D_A = <\!1-{{f_\nu(\lambda1050-1170)_{obs}}
\over {f_\nu(\lambda 1050-1170)_{pred.}}}\!>$ (Oke \&  Korycanski
1982), and is typically 0.58$\pm$0.09  in z$>$4 quasars (Schneider,
Schmidt \& Gunn 1991a,b) or radio galaxies (\eg, Spinrad, Dey \& Graham
1995).  Assuming an intrinsically flat continuum spectrum (\ie,
$f_\nu\propto\nu^0$) for \zg5a, we measure $D_A > 0.70$ (3$\sigma$
lower limit). This is consistent with the theoretical estimate of
$D_A(z=5.34) \approx 0.79$ of Madau (1995). For comparison, the two
next highest redshift objects, the $z=4.92$ lensed galaxy in the field
of CL~1358+62 and the $z=4.897$ QSO PC1247+3406, have $D_A$ measures of
$\simgt 0.6$ and $\approx 0.64$ respectively (Franx \etal 1997,
Schneider \etal 1991b).

\section {Discussion}

The luminosity of \zg5a\ is not likely to be dominated by an active
galactic nucleus (AGN). The \lya\ emission line is narrow (deconvolved
${\rm FWHM \approx 280~\kms}$), and is more typical of starburst galaxies
than luminous AGN. The rest-frame \lya\ equivalent width of $\approx
95\pm 15$\AA\ is well within the range expected for dust--free
star--forming galaxies (Charlot \& Fall 1993). Although the sharp blue
edge of the \lya\ line implies that the measured equivalent width and
line width are underestimates, the equivalent width would have to be
larger by more than a factor of two before the observed line strength
requires a nonstellar ionizing continuum.  Finally, the lack of strong
emission lines of \ion{N}{5}$\lambda\lambda$1239,1243 and
\ion{Si}{4}$\lambda\lambda$1394,1403 argues against an AGN as the
central ionizing source.  The 1$\sigma$ limit on the
\ion{N}{5}/\lya\ ratio is 0.01, whereas most Sy II galaxies have ratios
ranging from $0.03 - 0.3$ (Heckman, personal communication; Kriss \etal
1992). The \ion{N}{5}/\lya\ limit should be considered a firm upper
bound, since the \lya\ flux is affected by absorption.


If the ionization is dominated by the young, hot stars, the observed
flux of the \lya\ emission line may be used to estimate a lower bound
to the star--formation rate (\Mdot) in the galaxy. The \lya\ luminosity
corresponds to an equivalent of $\sim 10^4$ O5 stars. Using the
canonical Case~B \lya/H$\alpha$ ratio of $\approx 10$ (Osterbrock
1989), and the conversion from H$\alpha$ luminosity to \Mdot\
for a Salpeter IMF between $0.1<M<125\Msun$ (Madau, Pozzetti \&
Dickinson 1998; Kennicutt 1983) the $L_\lya$ implies $\Mdot\sim
6h_{50}^{-2}\Msun~yr^{-1}$ (\qnot=0.5; \Mdot\ is $\approx 3.3$
times larger for \qnot=0.1).  These are lower limits to \Mdot\ since we
have not made any correction to the \lya\ flux either for dust
extinction or absorption internal to \zg5a.

The star--formation rate may also be estimated from the observed UV
continuum emission.  If the UV continuum light in \zg5a\ is unreddened
and has a spectral slope of $f_\nu\propto \nu^0$, the implied
star--formation rate is $\approx 6\Msun~yr^{-1}$. Here, we have adopted the
conversion from $L_{1500}$ to \Mdot\ calculated by Madau, Pozzetti \&
Dickinson (1998) for a $>100$~Myr old population with a Salpeter IMF
($0.1<M<125\Msun$) of $\Mdot\approx10^{-40}\times L_{1500}$.
(We note that this is roughly consistent with the rate
derived from the Leitherer \& Heckman (1995) models, which is
calculated for a different IMF and much younger ages of $<10$~Myr).  These
conversions are uncertain (they depend on the assumed star--formation
history, IMF, metallicity and age) and are meant to be illustrative
rather than definitive.  It is noteworthy that the \Mdot\ implied by
$L_{\rm 1500}$ is very similar to that derived from $L_\lya$; in this
respect \zg5a\ differs from the bulk of the $z\sim 3$ starburst
population, which tend to have much smaller $L_\lya/L_{1500}$ (Steidel,
personal communication).  The agreement in \Mdot\ derived by these
different methods may suggest that the \lya\ and UV continuum emission
from \zg5a\ are not significantly attenuated by dust, or that the
geometry and kinematics of the interstellar medium permit most
\lya\ photons to escape.

\zg5a\ is spatially resolved in our ground-based $I$-band image
(deconvolved FWHM $\approx 0\arcsper7 \approx  3.9h_{50}^{-1}$~kpc for
\qnot=0.5, or $7h_{50}^{-1}$~kpc for \qnot=0.1). In comparison, the
bulk of the known luminous $z\sim 2.5-3.5$ starburst galaxies are
compact systems (half--light radii $\sim 0\arcsper2-0\arcsper3$) with
radial surface brightness profiles more similar to those of dynamically
relaxed elliptical and bulge-dominated galaxies (Giavalisco, Steidel \&
Macchetto 1996). The somewhat lower luminosity $z\sim 3$ starburst galaxies
observed by Lowenthal \etal (1997) show slightly larger half-light
radii ($\sim 0\arcsper2-0\arcsper6$) with varied morphologies, but
these are also compact systems.  This difference may suggest that
\zg5a\ is being observed in an earlier stage in its formation, or that
it is composed of multiple sub-clumps.  We note, however, that the
$I$-band filter contains the \lya\ emission line which may well dictate
the observed extended morphology in this band.  High spatial resolution
{\it HST} imaging of the continuum emission will be necessary to
further investigate the morphology of the galaxy.


What is the luminosity of \zg5a?  For comparison, a local $M(B)=-21$
Magellanic irregular star--forming galaxy placed at $z=5.34$ will have
$I\approx25.3$ (\hnot=50\hnotunit and \qnot=0.5; $I\approx26.6$ if
\qnot=0.1). Since \zg5a\ has a line-corrected $I$-band magnitude of
26.5, it is perhaps comparable or slightly less luminous. A more
pertinent comparison may be made to the known population of
star--forming galaxies at $z\sim 2.5-3.5$: the UV (1500\AA) luminosity
function of these galaxies is fairly well modelled by a Schechter
function with $M_{1500}^*=-21$~AB mag (for \hnot=50, \qnot=0.5;
Dickinson 1998).  For the same cosmology, the absolute magnitude of
\zg5a\ is $M_{1500}=-20$~AB mag, and is therefore sub-luminous in
comparison to an $L^*_{1500}$ $z\sim 3$ starburst galaxy.

A robust, dynamical estimate of the mass is difficult to obtain from
the current data, since the UV continuum contains emission from only
the most massive (\ie, youngest) stars, and the \lya\ emission line is
a poor dynamical probe since its width may be affected by resonance
scattering, winds, and absorption.  Nevertheless, under the assumption
that radiative transfer and non-gravitational kinematics act to only
broaden the line, the width of the emission line (${\rm FWHM \approx
280\kms}$) can provide an upper limit to the mass of the galaxy.
Assuming that the blue half of the \lya\ line is absorbed, the velocity
dispersion of the gas ($\sim 240\kms$) implies an upper limit of
$M_{dyn}< 2.6\times10^{10}h_{50}^{-1}$\Msun\ (\qnot=0.5). A lower limit
to the mass in the interstellar medium may be estimated from the mass
in the ionized component. If the \lya\ recombination radiation arises
from ionized gas which uniformly fills a spherical volume of radius
$\sim 2~h_{50}^{-1}$~kpc, the mass in the ionized medium is
$M_{ion}(\sim 10^4K) \sim 1.8\times 10^9 \Msun$ (\qnot=0.5).  
Although these constraints on the mass are poor and nothing is known
about the mass or luminosity function of galaxies at $z>5$, it is
noteworthy that both the mass and luminosity of \zg5a\ are
representative of an object less luminous than an $L^*_{1500}$ star--forming
galaxy at $z\sim 3$ (Dickinson 1998).


Is \zg5a\ a truly primeval object undergoing its first
episode of star--formation? The present data are of too low
signal-to-noise ratio to adequately address this question.  The only
way to do so would be to obtain high signal-to-noise ratio spectra of
the continuum light to place meaningful limits on the column density of
metals in the galaxy. Traditional models of galaxy formation which
hypothesize spheroid formation via a monolithic collapse predict that
most of the stars would be formed at a rapid rate (\eg, Eggen,
Lynden-Bell \& Sandage 1962, Meier 1976). This does not appear to be
the case for \zg5a:  the derived star--formation rate is not very high,
and is instead quite similar to that observed in sub-$L^*_{1500}$ $z\sim 3$
star--forming galaxies. This may imply that \zg5a\ is either simply a
higher $z$ counterpart of these galaxies --- an object which has
already undergone the bulk of its star--formation --- or that it is an
intrinsically low luminosity galaxy in the process of formation.

One caveat to this interpretation of the derived star--formation rate
is that we have not accounted for any extinction by dust internal to
\zg5a.  Meurer \etal (1997) have persuasively argued that dust
extinction in $z\sim 3$ starburst galaxies implies that the observed
fluxes (and therefore the star--formation rates) in these systems are
underestimated by as much as an order of magnitude.  However, the
strong \lya\ emission line and the agreement in the star--formation
rates derived using the emission line and the UV continuum emission,
together suggest that the observed UV spectrum of \zg5a\ is not
strongly reddened.

Near-IR observations may eventually constrain both the age and the
extinction of the starlight in the galaxy:  a young, star--forming
galaxy at $z=5.34$ will have colors of $I-K \approx 1.7$, $I-H \approx
1.1$ and $I-J \approx 0.6$.  If \zg5a\ is dust free, it will have
near-IR magnitudes of $J\approx 26.4$, $H\approx 25.9$, and $K\approx
25.4$.  Existing $K_S$ imaging of the field by van Breugel \etal (1998)
places a $1\sigma$ upper limit of 24.0~mag (in an 1\arcsper5 aperture) on
the continuum emission. This suggests an upper limit to the age of
$\simlt 200$~Myr (for a dust--free instantaneous burst solar
metallicity population --- no useful constraints can be derived from a
constant star--forming model; Bruzual and Charlot 1996), or
alternatively an upper limit to the extinction of $E_{B-V} < 0.4$ (for
a zero age starburst population extincted a foreground screen of dust
following the extinction law of Kinney \etal 1994). Deeper near-IR
measurements are necessary to provide better constraints.

The strong \lya\ emission from \zg5a\ suggests that ongoing deep
emission line imaging searches, although limited in spectral coverage,
are likely to be efficient ways of probing the high--$z$ Universe.
These surveys have already begun to yield promising candidates (\eg,
Cowie \& Hu 1998, Thommes \etal 1998) at comparable or brighter
emission line fluxes (perhaps representing similar or more actively
star--forming systems) which await spectroscopic confirmation.
Spectroscopic redshifts for galaxies at $z\simgt 5$ will be
increasingly difficult to obtain at optical wavelengths, given the
decreasing CCD response at long-wavelengths and the increasing sky
background.  The UV spectral lines from heavy elements may be extremely
weak and redshift confirmation even at near-IR wavelengths may be hopeless
(for $z>5.5$, [\ion{O}{2}]$\lambda$3726,3729 is at
$\lambda_{obs}>2.42\mu m$).  Moreover, since [\ion{O}{2}] line fluxes
are expected to be $<0.1$ of the \lya\ flux in dust free systems
(Thompson, Djorgovski \& Beckwith 1994), near-IR searches must probe
fluxes $<10^{-18}\ {\rm erg\ s^{-1}\ cm^{-2}}$ to detect objects
similar to \zg5a.  Spectroscopic redshifts at $z>5$ may be solely
contingent on the asymmetry of the emission line and detection of the
spectral break across it: for objects like \zg5a, this task tests the
limit of the current instrumentation on the largest available
ground-based telescopes.  There may be other more luminous $z>5$
galaxies lurking behind the OH veil, well within reach of the new
generation of large ground-based telescopes.  However, if \zg5a\ is the
typical sub-galactic building block in a hierarchical galaxy formation
scenario (\eg, Baron \& White 1987), our investigation of this
protogalactic population must await the spatial resolution, low
background, and high sensitivity of the Next Generation Space
Telescope.

\acknowledgements

We thank J. Aycock, W. Wack, R. Quick, T. Stickel, G.  Punawai, R.
Goodrich, R. Campbell, T. Bida and B. Schaeffer for their invaluable
assistance during our observing runs at the W.~M.~Keck Observatory.  We
are grateful to A. Phillips for providing software and
assistance in slitmask construction and alignment; to W.
van Breugel for his $K_S$ image of the field; and to M. Dickinson, B.
Jannuzi, T. Heckman and J. Najita for very useful comments on the
manuscript. The W.\ M.\ Keck Observatory is a scientific partnership
among the Univ. of California, the California Institute of Technology,
and NASA, and was made 
possible by the generous financial support of the W.\ M.\ Keck
Foundation.  H.\ S.\ acknowledges support from NSF grant AST 95-28536.
A.\ D.\ acknowledges the support of NASA HF-01089.01-97A and partial
support from a Postdoctoral Research Fellowship at NOAO, operated by
AURA, Inc. under cooperative agreement with the NSF.

\pagebreak

\centerline {\bf References}

\medskip

\refindent Baron, E.\ \& White, S.\ D.\ M.\ 1987, \apj, 322, 585

\refindent Burstein, D.\ \& Heiles, C.\ 1982, \aj, 87, 1165

\refindent Bruzual, G.-A.\ \& Charlot, S.\ 1996, GISSEL Population Synthesis Models

\refindent Cowie, L.\ L.\ \& Hu, E.\ M.\ 1998, \aj, in press;
astro-ph/9801003

\refindent Charlot, S.\ \& Fall, S.\ M.\ 1993, \apj, 415, 580

\refindent Dickinson, M.\ E.\ 1998, in proceedings of the STScI May
1997 Symposium ``The Hubble Deep Field,'' eds. M.\ Livio, S.\ M.\ Fall \&
P.\ Madau; astro-ph/9802064

\refindent Djorgovski, S., Pahre, M.\ A., Bechtold, J.\ \& Elston, R.\ 1996, 
\nature, 382, 234

\refindent Djorgovski, S., Strauss, M.\ A., Spinrad, H., McCarthy, P.\ \& 
Perley, R.\ A.\ 1987, \aj, 93, 1318

\refindent Eggen, O.\ J., Lynden-Bell, D.\ \& Sandage, A.\ 1962, \apj, 136, 748

\refindent Franx, M., Illingworth, G.\ D., Kelson, D.\ D., van Dokkum,
P.\ G.\ \& Tran, K.-V.\ 1997, \apj, 486, L75

\refindent Giavalisco, M., Steidel, C.\ \& Macchetto, F.\ D.\ 1996, \apj, 470, 189

\refindent Hamann, F.\ 1997, \apjsupp, 109, 279

\refindent Hu, E.\ M., McMahon, R.\ G.\ \& Egami, E.\ 1996, \apj, 459, L53

\refindent Kennicutt, R.\ C.\ 1983, \apj, 272, 54

\refindent Kennicutt, R.\ C.\ 1992, \apj, 388, 310

\refindent Kinney, A.\ L., Calzetti, D., Bica, E.\ \& Storchi-Bergmann,
T.\ 1994, \apj, 429, 172

\refindent Kriss, G.\ A., Davidsen, A., Blair, A.\ F., Ferguson,
H.\ C., \& Long, K.\ S.\ 1992, \apj, 394, L37

\refindent Leitherer, C.\ \& Heckman, T.\ M.\ 1995, \apjs, 96, 9

\refindent Lequeux, J., Kunth, D., Mas-Hesse, J.\ M.\ \& Sargent, W.\ L.\ W.\ 
1995, \aa, 301, 18

\refindent Liu, C.\ T.\ \& Kennicutt, R.\ C.\ 1995, \apj, 450, 547

\refindent  Lowenthal, J. D., Hogan, C. J., Green, R. F., Caulet, A.,
Woodgate, B. E., Brown, L., Foltz, C. B.\ 1991, \apj, 377, L73

\refindent Lowenthal, J.\ D., Koo, D.\ C., Guzm\'an, R., Gallego, J., 
Phillips, A.\ C., Faber, S.\ M., Vogt, N.\ P., Illingworth, G.\ D.\ \& 
Gronwall, C.\ 1997, \apj, 481, 673

\refindent Lu, L., Sargent, W.\ L.\ W., Barlow, T.\ A., Churchill, C.\ W.\ \& 
Vogt, S.\ S.\  1996, ApJS, 107, 475

\refindent Madau, P. 1995, \apj, 441, 18

\refindent Madau, P., Ferguson, H.\ C., Dickinson, M.\ E., Giavalisco, M., 
Steidel, C.\ C., \& Fruchter, A.\ 1996, \mnras, 283, 1388

\refindent Madau, P., Pozzetti, L.\ \& Dickinson, M.\ 1998, \apj, in press.

\refindent Massey, P.\ \& Gronwall, C.\ 1990, \apj, 358, 344

\refindent Massey, P., Strobel, K., Barnes, J.\ V.\ \& Anderson, E.\ 1988, 
\apj, 328, 315

\refindent Meier, D.\ L.\ 1976, \apj, 207, 343

\refindent Meurer, G.\ R., Heckman, T.\ M., Lehnert, M.\ D., Leitherer,
C., \& Lowenthal, J.\ 1997, \aj, 114, 54

\refindent M{\o}ller, P.\ \& Warren, S.\ J.\ 1993, \aa, 270, 43

\refindent Oke, J.\ B., Cohen, J.\ G., Carr, M., Cromer, J., Dingizian, A.,
Harris, F.\ H., Labrecque, S., Lucino, R., Schaal, W., Epps, H., \& Miller,
J.\ 1995, \pasp, 107, 375.

\refindent Oke, J.B. \& Korycanski, D.G. 1982, \apj, 255, 11

\refindent Osterbrock, D.\ E.\ 1989, Astrophysics of Gaseous Nebulae and Active
Galactic Nuclei (University Science Books: California)

\refindent Pascarelle, S.\ M., Windhorst, R.\ A., Driver, S.\ P., Ostrander, 
E.\ J.\ \& Keel, W.\ C.\ 1996, \apj, 456, L21

\refindent Pritchet, C.\ J.\ 1994, \pasp, 106, 1052

\refindent Pritchet, C.\ J.\ \& Hartwick, F.\ D.\ A.\ 1990, \apj, 355, L11

\refindent Rawlings, S., Lacy, M., Blundell, K.\ M., Eales, S.\ A.,
Bunker, A.\ J., \& Garrington, S.\ T.\ 1996, Nature, 383, 502

\refindent Schneider, D.\ P., Schmidt, M.\ \& Gunn , J.\ E.\  1991$a$, \aj, 101, 2004 

\refindent Schneider, D.\ P., Schmidt, M.\ \& Gunn , J.\ E.\  1991$b$, \aj, 102, 837 

\refindent Spinrad, H., Dey, A., \& Graham, J.\ R.\ 1995, \apj, 438, L51

\refindent Spinrad, H., Dey, A., Stern, D., Dunlop, J., Peacock, J.,
Jimenez, R., \& Windhorst, R.\ 1997, \apj, 484, 581

\refindent Stanford, S.\ A., Eisenhardt, P.\ R.\ M.\ \& Dickinson, M.\ 1995, \apj, 450, 512

\refindent Steidel, C.\ C.\ \& Hamilton, D.\ J.\ 1992, \aj, 104, 941

\refindent Steidel, C.\ C., Giavalisco, M., Pettini, M., Dickinson, M., \& 
Adeleberger, K.\ L.\ 1996, \apj, 462, L17

\refindent Thommes, E., Meisenheimer, K., Fockenbrock, R., Hippelein, H., 
R\"oser, H.-J., \& Beckwith, S.\ 1998, \mnras, 293, L6

\refindent Thompson, D.\ J., Djorgovski, S.\ \& Beckwith, S.\ 1994, \aj, 
107, 1

\refindent van Breugel, W.\ \etal 1998, submitted to \aj


\begin{deluxetable}{cccccccccc}
\leftmargin 0.0in 
\small
\tablewidth{6.9in}
\tablehead{
\colhead{Name} 
& \colhead{Redshift} 
& \colhead{$I$\tablenotemark{\dag}} 
& \colhead{\it B$-$I\tablenotemark{\dag}} 
& \colhead{\it R$-$I\tablenotemark{\dag}} 
& \colhead{$F_{\rm Ly\alpha}$\tablenotemark{\dag\dag}} 
& \colhead{$W_{\rm Ly\alpha}^{\rm obs}$\tablenotemark{\ddag}}
& \colhead{$\langle F_{1265}\rangle$\tablenotemark{\S}} 
& \colhead{$\langle F_{1500}\rangle$\tablenotemark{\S}} 
}
\startdata
 0140+326~BD3 &  4.022$\pm$0.002 & 25.1 & $>$3.9 &  0.2 & 5.4$\pm$0.1& 700$\pm$100 & $12.5\pm0.7$ & $11.0\pm0.8$ \nl
 \zg5a &   5.348$\pm$0.002\tablenotemark{\star} &  26.1 & $>$2.9 & $>$1.7  & 3.5$\pm$0.1 & 600$\pm$100 & $5.3\pm0.8$ & \nl
\enddata
\tablenotetext{\dag}{Vega--based magnitudes measured in
1\arcsper5 diameter apertures. Magnitudes presented in
this table have not been corrected for Galactic extinction and
reddening. Limits quoted are 1$\sigma$ upper limits.}
\tablenotetext{\dag\dag}{Line fluxes are in units of $10^{-17}\ {\rm erg\ s^{-1}\ cm^{-2}}$, and have been corrected for Galactic extinction using 
$E_{B-V}=0.035$.}
\tablenotetext{\ddag}{Observer's frame equivalent widths in \AA\ are 
estimated using a Gaussian fit to the emission line and linear fit
to the continuum redward of the emission line.}
\tablenotetext{\star}{The redshift is based on a Gaussian fit to the
peak of the \lya\ emission line. This may be slightly redward of the
true systemic velocity of the galaxy.}
\tablenotetext{\S}{$F_\lambda$ is in units of ${\rm
10^{-20}\ erg\ s^{-1}\ cm^{-2}\ \AA^{-1}}$.  $\langle
F_{1265}\rangle$ and $\langle F_{1500}\rangle$
are measured between the observed wavelengths corresponding to
1220$-$1310\AA\ and 1450$-$1550\AA\ respectively, from spectra
corrected for Galactic extinction of $E_{B-V}=0.035$. There is no continuum 
detection at $\sim 1500$~\AA\ for \zg5a.}
\label{tab1}
\end{deluxetable}

\begin{deluxetable}{ccccc}
\tablewidth{0pt}
\tablehead{
\colhead{Name} & 
\colhead{$L_{\lya}h_{50}^2(\qnot=0.5)$\tablenotemark{\dag}} &
\colhead{$L_{1500} h_{50}^2$(\qnot=0.5)\tablenotemark{\dag}} & 
\colhead{${\rm SFR_\lya} h_{50}^2$\tablenotemark{\ddag}} &
\colhead{${\rm SFR_{1500}} h_{50}^2$\tablenotemark{\ddag}} 
}
\startdata
 0140+326~BD3 & $7.2\times 10^{42}$ & $7.3\times 10^{40}$ & 5 & 7 \nl 
 0140+326~RD1 & $8.8\times 10^{42}$ & $6.0\times 10^{40}$\tablenotemark{\S} & 6 & 6 \nl 
\enddata
\tablenotetext{\dag}{Not corrected for reddening intrinsic to the
galaxies. $L_\lya$ and $L_{1500}$ are in units of ${\rm erg\ s^{-1}}$
and ${\rm erg\ s^{-1}\ \AA^{-1}}$ respectively.}
\tablenotetext{\ddag}{Star-formation rates (in units of
\Msun\ $yr^{-1}$) assume a Salpeter IMF with $0.1<M<125\Msun$ (see Madau, 
Pozzetti \& Dickinson 1998).  For
\qnot=0.1, these rates are $\sim$2.7 and 3.3 times larger for BD3 and
RD1 respectively.}
\tablenotetext{\S}{Estimated from the continuum flux at 1265\AA\ assuming 
a $F_\lambda\propto\lambda^{-2}$ spectrum.}
\label{bdrops2}
\end{deluxetable}


%
%

\end{document}